\documentstyle[pra,aps,amsfonts,twocolumn,epsf]{revtex}
\newcommand{\trace}{\mathop{\rm Tr}\nolimits}
\newcommand{\bra}[1]{\langle#1|}
\newcommand{\ket}[1]{|#1\rangle}

\begin{document}
\draft
\title{Negativity and Concurrence of mixed $2\times2$ states}
\author{Koenraad Audenaert\cite{KAmail},
Frank Verstraete\cite{FVMail},
Tijl De Bie\cite{TDBmail},
Bart De Moor\cite{BDMmail}}
\address{Katholieke Universiteit Leuven, \\
Dept. of Electrical Engineering (ESAT), Research Group SISTA \\
Kardinaal Mercierlaan 94, B-3001 Leuven, Belgium
}
\date{\today}
\maketitle
\begin{abstract}
We consider two measures of entanglement of mixed bipartite states of
dimension $2\times2$: concurrence and negativity. We first prove the conjecture
of Eisert and Plenio that concurrence can never be smaller than negativity. We then 
characterise all states for which concurrence equals negativity and also
those states for which the difference between concurrence and negativity is 
maximal (keeping either the concurrence fixed, or the participation ration
$R=1/\trace \rho^2$).
\end{abstract}
%------------------------------------------------------------------ BODY
In this Letter we investigate in an algebraic fashion the link between
two measures of entanglement for $2\times2$ quantum systems, 
known as concurrence and negativity, respectively.
As defined by Wootters \cite{wootters97} the concurrence of a mixed state $\rho$
is given by:
\begin{equation}
C(\rho) = \max(0,\sigma_1-\sigma_2-\sigma_3-\sigma_4),
\end{equation}
where the non-negative numbers $\sigma_i$ are the singular values (by convention
sorted in descending fashion) of the matrix
\begin{equation}
Q=\sqrt{\rho}^T \sigma_y\otimes\sigma_y \sqrt{\rho},
\end{equation}
$\sigma_y$ is the well-known Pauli matrix, and $\sqrt{\rho}$ is any matrix
satisfying 
\begin{equation}
\rho = \sqrt{\rho}\sqrt{\rho}^\dagger.
\end{equation}
The importance of this measure follows from the direct connection between concurrence
and entanglement of formation $E_f$: 
\begin{equation}
E_f(\rho) = -\mu_1\log_2\mu_1-\mu_2\log_2\mu_2,
\end{equation}
where
\begin{equation}
\mu_{1,2} = (1\pm\sqrt{1-C(\rho)^2})/2.
\end{equation}
One can prove \cite{wootters97} that $\rho$ is separable if and only if the concurrence
is zero.

Another test for separability of $2\times2$ systems is the Peres criterion \cite{peres96},
which states that $\rho$ is separable iff its partial transpose is positive (semi-)definite.
Since we are dealing with bipartite states, the row- and column-indices of the density matrix
$\rho$ can be split up in subindices pertaining to each of the two subsystems.
In this Letter we will use the notation $\rho_{(ii'),(jj')}$, unprimed and primed
indices referring to subsystem A and B, respectively.
The subsystem-B partial transpose $\rho^{T_B}$ corresponds to transposing primed indices only.
In general, this operation does not retain the positive definiteness required from any
density matrix, unless, as proven by Horodecki 
\cite{horodecki96a}, the state is separable. By definition, a Hermitian matrix is positive 
(semi-)definite iff its eigenvalues are positive (non-negative). In this respect, the
smallest eigenvalue of a Hermitian matrix can be considered as a kind of measure describing
the deviation from positive definiteness, or, applied to the partial transpose of a state,
as a measure of entanglement of the state.
More precisely, one defines the {\em negativity} $E_N$ of a state as:
$$
E_N(\rho) = \max(0,-2 \lambda_1(\rho^{T_B})),
$$
where $\lambda_i(X)$ are the eigenvalues of a Hermitian matrix $X$, by convention sorted
in {\em ascending} order.
The factor 2 in this definition is added for convenience.

For pure states, it has been proven that the negativity is exactly equal
to the concurrence \cite{vidal}. For mixed states, Eisert and Plenio conjectured that negativity
never exceeds concurrence (\cite{eisert}; question (ii) in \cite{zycz99}). 
In this Letter, we first give a proof of this conjecture and then go on to
characterize the mixed states for which equality between $C$ and $E_N$ holds.
In \cite{zycz99} the question was posed what the maximal difference between
$C$ and $E_N$ must be, in function of the participation ratio $R=1/\trace{\rho^2}$ (question (iii)).
This maximality case turns out to be much more difficult to treat than the equality case.
We first consider the class of so-called maximally entangled states \cite{fv1,ishi}. For
these states a simple analytic expression exists
for concurrence as well as for negativity, so that it is easy to calculate the maximal value
of $C-E_N$ within this class. Finally, we have conducted a numerical search for states with maximal
$C-E_N$. 

In order to establish a number of necessary relations, we first rederive Vidal and Tarrach's result
that $C=E_N$ for pure states.
A state vector $\psi_{(ii')}$ of a bipartite state can be reshaped to matrix form by interpreting
the unprimed index as a row index, and the primed index as a column index. Denoting
the matrix thus obtained by $\tilde{\psi}$, we have $\tilde{\psi}_{ii'} = \psi_{(ii')}.$
Every matrix has a singular value decomposition (SVD) $A=U\Sigma V^\dagger$, where $U$ and $V$ are
unitary and $\Sigma$ is a diagonal matrix with non-negative diagonal elements, called the
singular values. Applied to the reshaped  state vector, we get the Schmidt-decomposition of
the vector. The Schmidt-coefficients are just the square roots of the singular values.
In the $2\times2$ case, and for a normalised vector, there are two singular values, $\sigma_1$ and
$\sigma_2$ and $\sigma_1^2+\sigma_2^2=1$. By convention, $U$ and $V$ are chosen so that $\sigma_1\ge\sigma_2$.

Using these notations, the partial transpose of a pure state can be easily expressed.
\begin{eqnarray*}
(\rho^{T_B})_{(ii'),(jj')} &=& \rho_{(ij'),(ji')} \\
&=& \psi_{(ij')}\psi_{(ji')}^* \\
&=& \tilde{\psi}_{ij'} \tilde{\psi}^\dagger_{i'j} \\
&=& (\tilde{\psi} \otimes \tilde{\psi}^\dagger)_{(ii'),(j'j)};
\end{eqnarray*}
introducing the matrix $P_0=\sum_{ij} e^{ij} \otimes e^{ji}$, where
$e^{ij}$ is the standard matrix basis element containing just a single 1
on row $i$, column $j$, we can rewrite this concisely as
\begin{equation}
\rho^{T_B} = \tilde{\psi} \otimes \tilde{\psi}^\dagger P_0.
\end{equation}
Inserting now the SVD of $\tilde{\psi}=U\Sigma V^\dagger$ and using the product reversal 
property of $P_0$ (\cite{horn91}, 4.3.10):
\begin{eqnarray*}
\rho^{T_B} &=& (U\otimes V)(\Sigma\otimes \Sigma)(V^\dagger\otimes U^\dagger) P_0 \\
&=&(U\otimes V)(\Sigma\otimes \Sigma)P_0(U\otimes V)^\dagger.
\end{eqnarray*}
The explicit form of the matrix $(\Sigma\otimes \Sigma)P_0$
is
\begin{equation}
\left(
\begin{array}{cccc}
\sigma_1^2 & 0 & 0 & 0 \\
0 & 0 & \sigma_1\sigma_2 & 0 \\
0 & \sigma_1\sigma_2 & 0 & 0 \\
0 & 0 & 0 & \sigma_2^2
\end{array}
\right)
\end{equation}
and its eigenvalue decomposition (EVD) equals $W\Lambda W^\dagger$, where
\begin{equation}
W=\left(
\begin{array}{cccc}
1 & 0 & 0 & 0 \\
0 & 1/\sqrt{2} & 1/\sqrt{2} & 0 \\
0 & 1/\sqrt{2} & -1/\sqrt{2} & 0 \\
0 & 0 & 0 & 1
\end{array}
\right)
\end{equation}
and
\begin{equation}
\Lambda=\left(
\begin{array}{cccc}
\sigma_1^2 & 0 & 0 & 0 \\
0 & \sigma_1\sigma_2 & 0 & 0 \\
0 & 0 & -\sigma_1\sigma_2 & 0 \\
0 & 0 & 0 & \sigma_2^2
\end{array}
\right)
\end{equation}
The negativity of the pure state is, therefore, explicitly given by
\begin{equation}
E_N(\ket{\psi}\bra{\psi}) = 2\sigma_1\sigma_2 = 2|\det\tilde{\psi}|.
\end{equation}

In order to calculate the concurrence of the pure state, we first need 
the square root $\sqrt{\rho}$; in this case, this is just the state vector
$\psi$. The matrix $Q$ is therefore a scalar, equal to
\begin{eqnarray*}
Q &=& \psi^T \sigma_y\otimes\sigma_y \psi \\
&=& \psi_1 \psi_4-\psi_2\psi_3-\psi_3\psi_2+\psi_4\psi_1 \\
&=& 2\det\tilde{\psi},
\end{eqnarray*}
and has just a single singular value, equal to its own absolute value.
The concurrence is thus 
\begin{equation}
C=|Q|=2|\det\tilde{\psi}|=E_N.
\end{equation}
This concludes the proof.

We now turn to the first main result of this Letter, being the proof of
Eisert and Penio's conjecture that $C\ge E_N$ for mixed states. To that purpose
we need a characterization of concurrence for mixed states that has been presented
in \cite{wootters97}, though not in an explicit way:
the concurrence of a mixed state is the minimal average concurrence of the pure states
in any ensemble realising the state. In fact, this is analogous to the
definition of entanglement of formation for mixed states.
So, we have:
\begin{equation} \label{eq:cmix}
C(\rho) = \min_{p_i,\phi^i} \sum_i p_i C(\ket{\phi^i}\bra{\phi^i}),
\end{equation}
where the minimisation is over all ensembles $\{p_i,\phi^i\}$ for which
$\rho = \sum_i p_i \ket{\phi^i}\bra{\phi^i}$. Now, due to the equality $C=E_N$ for
pure states, we can replace $C$ by $E_N$ in (\ref{eq:cmix}). Furthermore,
replacing the negativities by their definition we get
\begin{equation}
C(\rho) = \min_{p_i,\phi^i} \sum_i p_i (-2)\lambda_1(\ket{\phi^i}\bra{\phi^i}^{T_B}).
\end{equation}
In principle, this should be an inequality, because some members of the ensemble could have
zero negativity (positive $\lambda_1$).
However, if $\rho$ is not separable, we can always write the equality sign, provided we restrict
ourselves to optimal ensembles in which every member has the same concurrence (such ensembles always exist
\cite{wootters97}).

A well-known inequality of Weyl states that the minimal eigenvalue of a sum of Hermitian matrices
is never smaller than the sum of the respective minimal eigenvalues:
\begin{equation}
\lambda_1(A+B) \ge \lambda_1(A)+\lambda_1(B).
\end{equation}
Hence,
\begin{eqnarray*}
C(\rho) &\ge& \min_{p_i,\phi^i} (-2) \sum_i \lambda_1(p_i\ket{\phi^i}\bra{\phi^i}^{T_B}) \\
&\ge& (-2) \lambda_1(\sum_i p_i\ket{\phi^i}\bra{\phi^i}^{T_B}) \\
&=& -2 \lambda_1(\rho^{T_B}),
\end{eqnarray*}
where the minimization has been dropped for obvious reasons.
Now, if $\rho$ is not separable, its negativity is non-zero, so that we finally obtain $C\ge E_N$.
If $\rho$ is separable, then both the negativity and the concurrence are zero, so that $C=E_N$.
This concludes the proof.

We now give a characterization of the states that have concurrence equal to negativity.
Trivially, separable states and pure states belong to this class. In general, the essential reason
why concurrence is not always equal to negativity is given by Weyl's inequality, as can be seen from the
previous proof. It is easy to see that Weyl's inequality becomes an equality if and only if
the eigenvectors of A and B pertaining to the respective smallest eigenvalues are equal (up to, possibly,
a scalar phase factor).
In the present case, these eigenvectors are
\begin{equation} \label{eq:eveq}
U^i\otimes V^i (0, 1, -1, 0)^T/\sqrt{2},
\end{equation}
where the pure states in the given ensemble for $\rho$ have an SVD 
$\tilde{\phi^i}=U^i\Sigma^i V^{i\dagger}$. We are allowed to restrict $U^i$ and $V^i$ to SU(2), because
the global phase of $\phi^i$ is not relevant.
If all these eigenvectors are equal, then equality holds in Weyl's inequality,
and the average concurrence of the ensemble equals the negativity of $\rho$.
But this is the lowest value the average concurrence of any ensemble realising $\rho$
can have (due to the previous theorem), so this ensemble must {\em automatically} be optimal
and its average concurrence must be equal to the concurrence!
In other words, $C=E_N$ if and only if the abovementioned eigenvectors of pure states in a realising
ensemble are equal. 

This result directly leads to a method for generating all mixed states with $C=E_N$.
The condition (\ref{eq:eveq}) can be rewritten (using vector-to-matrix reshaping) as
\begin{equation}
U^i \left(\begin{array}{cc}0 & -1 \\ 1 & 0\end{array}\right) V^{iT} =
U^1 \left(\begin{array}{cc}0 & -1 \\ 1 & 0\end{array}\right) V^{1T}.
\end{equation}
Using the property of SU(2)-matrices that $U\sigma_y=\sigma_y U^*$, this can be simplified to the condition
\begin{equation} \label{eq:eveq2}
V^{1\dagger}V^i = U^{1\dagger}U^i.
\end{equation}
By applying local unitary operations to the state $\rho$, $U^1$ and $V^1$ can be made equal to the unit matrix.
The condition then becomes $U^i=V^i$, and $\tilde{\phi^i} = U^i\Sigma U^{i\dagger}$. This condition therefore
amounts to imposing that $\tilde{\phi^i}$ is a positive semidefinite Hermitean matrix (PSDH).
Hence, a state $\rho$ has $C=E_N$ if and only if it there are local unitary operations $U$ and $V$ such that
$(U\otimes V)\rho(U\otimes V)^\dagger$ is in the convex closure of the set of pure states 
$\ket{\phi}\bra{\phi}$ with $\tilde{\phi}$ PSDH.

When attempting to use similar methods for the characterization of the states for which $C-E_N$ is {\em maximal} 
(w.r.t.\ some partitioning of the set of states), one immediately runs into the problem that the concurrence
is given as the result of a minimisation. While this minimisation vanishes automagically during the determination of
states with minimal $C-E_N$, this does not happen when maximizing $C-E_N$. Furthermore, while there is a Weyl's
inequality $\lambda_1(A+B)\le \lambda_1(A)+\lambda_n(B)$ ($n$ the matrix dimension), it is not necessarily so that
equality holds if and only if the eigenvector of $A$ pertaining to $\lambda_1(A)$ is equal to the eigenvector of $B$
pertaining to $\lambda_n(B)$; and even if this where the case, this result would not be applicable to the problem
at hand, as the latter eigenvector is a product vector ($U^i\otimes V^i (1,0,0,0)^T$) while the former is not 
($U^i\otimes V^i (0,1,-1,0)^T$).

In a previous paper \cite{fv1,ishi} we have characterized the states with maximal concurrence and maximal negativity,
respectively, w.r.t.\ the partitioning of the set of states in subsets of states with equal eigenvalues.
It turned out that the so-called maximally entangled (ME) states have both maximal $C$ and maximal $E_N$;
these values are given by
\begin{eqnarray*}
C &=& \max(0,\lambda_1-\lambda_3-2\sqrt{\lambda_2 \lambda_4}) \\
E_N &=& \max(0,\sqrt{(\lambda_1-\lambda_3)^2+(\lambda_2-\lambda_4)^2}-\lambda_2-\lambda_4),
\end{eqnarray*}
where $\lambda_i$ are the eigenvalues of the state (in contrast to our previous convention, now sorted in 
descending order). For the complete
expressions for the states themselves, we refer to \cite{fv1}. For these ME states it is easy
to calculate the maximal $C-E_N$ in function of, say, the participation ratio $R=1/\trace{\rho^2}=1/\sum_i\lambda_i^2$,
or of $C$ itself. 

Pure ME states have $R=1$ and $C=E_N=1$. For rank-2 states, $R=1/(\lambda_1^2+(1-\lambda_1)^2)$ and lies in the 
interval $[1,2]$, and $C=\lambda_1$ and $E_N=\sqrt{\lambda_1^2+(1-\lambda_1)^2}-(1-\lambda_1)$. After some basic algebra
we get $(C-E_N)_{\max}=1-1/\sqrt{R}$, the maximum of which is reached for $R=2$.

In the rank-3 case the calculations are somewhat more cumbersome. The result is that $C-E_N$ is maximal when
either $\lambda_1=\lambda_2$ ($R$ between 2 and 3) or $\lambda_3=0$ ($R$ between 1 and 2). 
The latter case corresponds to the rank-2 states, so in this interval rank-3 states are not optimal.
In the former case $(C-E_N)_{\max}$ is given by $(1+2\alpha-\sqrt{\alpha-4+15/R})/3$ with
$\alpha=\sqrt{-2+6/R}$. Finally, calculations reveal that rank-4 states are suboptimal in the interval $1\le R\le3$;
for $R$ between 3 and 4, the states are separable \cite{zycz2} so that $C-E_N=0$.
 
These results are depicted in Figure 1. Comparing this figure with the one in \cite{zycz99}, obtained using Monte-Carlo
calculations, we see that optimal ME states have a fairly large $(C-E_N)_{\max}$. In addition to these analytic
investigations, we have also performed numerical calculations to determine the actual maximum value of $(C-E_N)_{\max}$.
The calculations are based on the ``downhill simplex'' optimization method, available in the MatLab software package.
To avoid local optima, the calculations have been performed a large number of times for each $R$-value.
The resuts of this optimization suggest that, in the $R$-interval $[1,2[$, the optimal values of $C-E_N$ are larger
than can be obtained with optimal (rank-2) ME states;
the optimal states are also rank-2 states, however. These optimal $C-E_N$ values are drawn as dotted lines in Figure 1.
For $R$ in the interval $[2,3]$, the states that are optimal w.r.t.\ $C-E_N$ within the set of ME states seem to be
optimal within the set of all states too, which is somewhat unexpected. The case of $R$ larger than 3 is not depicted,
as all such states are separable and have $C-E_N$ equal to zero.
\begin{figure}[h]
\leavevmode
\epsfxsize=8cm
\epsfbox{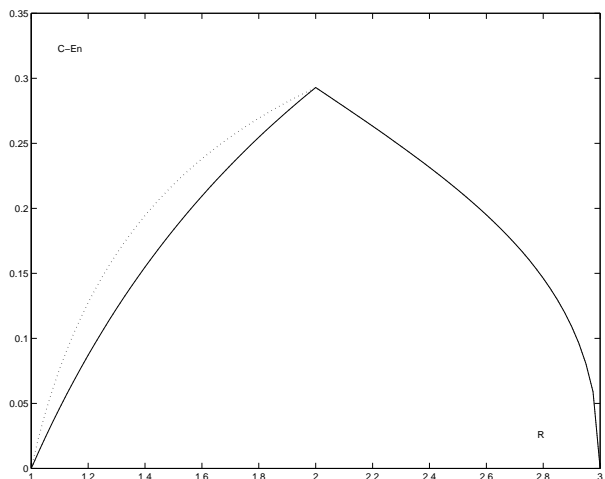}
\caption{Maximal difference of concurrence with negativity $C-E_N$ versus participation ratio $R$.
The full line represents analytical results obtained with maximally entangled states. The dotted line
represents numerical results obtained using downhill simplex optimization over the set of all possible states.}
\end{figure}

We can proceed in a similar way to find the states with maximal $C-E_N$ w.r.t.\ $C$ (instead of w.r.t.\ $R$).
Restricting ourselves again to maximally entangled states, it turns out that rank-2 ME states are again optimal within
the ME states.
Moreover, in contrast to the previous problem (maximal $C-E_N$ versus $R$), these ME-optimal states are also 
{\em the} optimal states in general, i.e.\ over all possible states \cite{intern}.
To prove optimality within the class of ME states, we first note that
negativity as a function of the eigenvalues $\lambda_i$ has no minimum over the finite positive reals,
so the actual minimum must occur on the boundary of the allowable set (simplex): $\lambda_i\ge 0$, $\sum_i\lambda_i=1$,
and $C$ prescribed. Calculations reveal that this minimum occurs for $\lambda_3=\lambda_4=0$. From the prescription 
for $C$ then follows that $\lambda_1=\max(C,1-C)$, $\lambda_2=\min(C,1-C)$ and $C-E_N=1-\sqrt{C^2+(1-C)^2}$ 
(see Figure 2).
\begin{figure}[h]
\leavevmode
\epsfxsize=8cm
\epsfbox{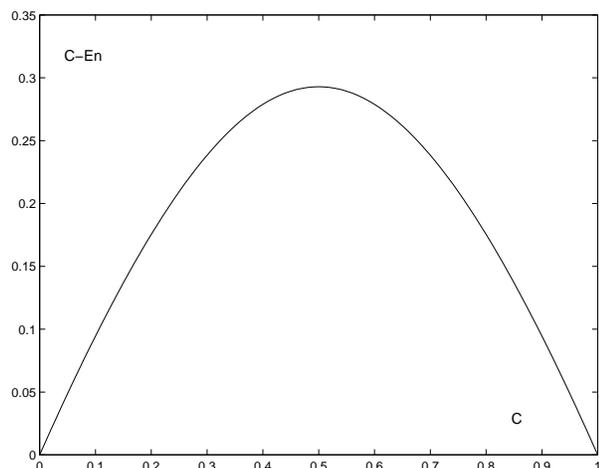}
\caption{Maximal difference of concurrence with negativity $C-E_N$ versus $C$.}
\end{figure}

Koenraad Audenaert is postdoctoral researcher with the K.U.Leuven. 
Frank Verstraete and Tijl De Bie are PhD students with the K.U.Leuven.
Bart De Moor is Full Professor at the K.U.Leuven.
This work is supported by several institutions: 
the Flemish Government
(Research Council K.U.Leuven: Concerted Research Action Mefisto-666;
FWO projects G.0240.99, G.0256.97, and Research Communities	ICCoS and ANMMM;
IWT projects EUREKA 2063-IMPACT and STWW),
the Belgian State
(IUAP P4-02 and IUAP P4-24; Sustainable Mobility Programme - Project MD/01/24),
the European Commission
(TMR Networks: ALAPEDES and System Identification;
Brite/Euram Thematic Network NICONET) and
Industrial Contract Research (ISMC, Electrabel, Laborelec, Verhaert, Europay)
The scientific responsibility is assumed by the authors.
%------------------------------------------------------------- BIBLIOGRAPHY

\end{document}